# Predicting one type of technological motion?
# A nonlinear map to study the 'sailing-ship' effect

Giovanni Filatrella
*INFN Gruppo Collegato Salerno and*
*Department of Science and Technology, University of Sannio,*
*Via Francesco de Sanctis sn, 82100 Benevento, Italy*
e-mail: filatrella@unisannio.it

Nicola De Liso
*Department of Law – Economics Division, University of Salento*
*Via per Monteroni sn, Building R1, 73100 Lecce, Italy*
e-mail: nicola.deliso@unisalento.it

**Abstract**

In this work we use a proven model to study a dynamic duopolistic competition between an old and a new technology which, through improved technical performance – e.g. data transmission capacity – fight in order to conquer market share. The process whereby an old technology fights a new one off through own improvements has been named 'sailing-ship effect'. In the simulations proposed, intentional improvements of both the old and the new technology are affected by the values of three key parameters: one scientific-technological, one purely technological and the third purely economic. The interaction between these components gives rise to different outcomes in terms of prevalence of one technology over the other.

*Keywords*: Nonlinear Dynamics; Sailing-ship effect; Econophysics

# Predicting one type of technological motion?
# A nonlinear map to study the 'sailing-ship' effect

**1. Introduction**

The importance of technology and its progress hardly needs to be emphasised, as it affects the daily life of us all. Within technology many dimensions – scientific, economic, political, juridical and 'pure' technological – interact and feed each other. A systematic fundamental interaction, though, is that between technology and the economy. Economic variables such as the rate of interest on capital needed for new investments or the willingness to reduce costs to face competition are fundamental. Economists have been interested in defining and measuring technical change and innovation since a long time, and three fruitful schools of thought, which are concerned with technical change, are the Classical, Neo-Classical and Evolutionary ones. Each school is characterised by some basic assumptions – a paradigmatic nucleus – and a set of tools. Each approach has become heavily mathematicised, making use of the most advanced mathematical and statistical techniques, ranging from the mathematics of chaos to the various generations of (G)ARCH models. The main point about technical change and innovation is that they are intrinsically dynamic phenomena (Walrave and Raven, 2016), so that the technical tools with which we analyse them must be adequate to the treatment of a dynamics which, often, is also non-linear and stochastic (Valenti et al., 2018).

Innovation is a particularly strong phenomenon in capitalist economies in which firms, hunting for profit, look for better ways of doing things, be it a product, process, or organisational innovation. Worthy of a comment here is the work of the social scientist Joseph A. Schumpeter (1911, 1950) who, through his emphasis on innovation in capitalistic economies – innovation being sought for the profit sake –, opened a new research path which from the early 1980s is addressed as *evolutionary economics*. Economic evolution takes place through a process of "creative destruction", of which innovation is the prime mover. The stress on innovation lies in that the innovative firm will often be for some time a monopolist, that is it will harvest *monopolistic* profits, i.e. profit margins unthinkable in competitive environments. Monopolistic profits, though, are transitory because of – among other things – imitation of other firms or the discovery by a new firm of yet another better way of doing the same thing(s).

An important phenomenon which is sometimes observed within our economies is the technological competition which develops between an incumbent ('old') technology and a new emerging technology aimed at providing similar services. The phenomenon according to which an old technology is *intentionally* improved *as a consequence* of the emergence of a new, potentially superior one, takes the name 'sailing-ship effect' (De Liso and Filatrella, 2008; Adner and Snow, 2010; Sick et al., 2016). This phrase synthesises the accelerated improvements experienced by old sail-ships as new steam-ships began to appear in the 1820s. Examples thus range from sail vs. steam in ships' construction (Gilfillan, 1935) to cables-cum-modems vs. fiber optics in data transmission (GTS, 2019). Furthermore, as a further collateral instance, in the pharmaceutical sector scientific investigation and technological development are normal when drugs, diagnostic or medical procedures are already adopted (Gagliardi *et al*., 2018). The question of market forecasting is tightly connected to the performances of the technologies – for instance to foresee the most effective technology in curbing greenhouse gases (Berggren *et al.*, 2009), where the uncertainty in the outcome of the R&D has led to a complex structure of alliances and competition (Borgstedt *et al.*, 2016).

Technological and economic considerations affect the process of innovation according to the sailing-ship principle.

In the first formalisation of the phenomenon (De Liso and Filatrella, 2008) the authors consider a dynamic duopolistic model in terms of a nonlinear functional map. In this model, which takes as the starting point the fundamental economic principle of profit maximisation, the old and the new technology are improved on the basis of the resources devoted to R&D, the efficiency according to which those resources are transformed into improved performance and the role played by the interest rate. The process of improvement of a technology is limited by the highest performance that can be reached by the technology itself: this is a physical limit, which usually sees the new technology being characterised by a higher potential. In the 2008 paper – as well as in a refined one in which a further 'memory effect' favouring the old technology is considered (De Liso and Filatrella, 2011) – the authors are content with showing the presence of the sailing-ship effect providing just one set of parameters.

It is however interesting to study – and this is what we do in this paper – what happens in the original model when different values of three key parameters, namely the efficiency in converting R&D into improved performance, the highest possible technical performance of both technologies and the interest rate, are systematically varied. The reason to focus on these parameters lies in that the first is scientific-technological in kind, the second is purely technological, while the third is eminently economic. Furthermore, when introducing the original model, we will make explicit some mathematical aspects which were not made explicit in the original model itself.

The work is structured as follows. In section 2 we recall some methodological aspects which connect physics with economics, and the application of these methods to a peculiar form of technical change. In section 3 we provide a résumé of the first formal model concerned with the 'sailing-ship effect'. In section 4 we use that model to study the performance of the two technologies when the key parameters take different values. In section 5 the conclusions are drawn.

**2. An 'econophysics' approach to technical change**

Many research fields share the aim of predicting the motion of their subject matter – be it the position of planets, the weather or financial markets – and mathematics and statistics are often powerful tools which actually allow one to predict the evolution of certain phenomena. Often the same methods are used in different areas on the basis of analogies which are supposed to justify this use.

Through analogies one can find new and different ways to use existing tools such as mathematics. Oppenheimer writes that analogy is an indispensable and inevitable tool for scientific progress, clarifying that by 'analogy' he means:

"a special kind of similarity, which is the similarity of structure, the similarity of form, a similarity of constellation between two sets of structures, two sets of particulars, that are manifestly very different but have structural parallels. It has to do with relation and interconnection." (Oppenheimer, 1956, p. 129)

In the economics discipline, starting with the Marginalist-Neoclassical revolution of the 1870s, and in particular with the works of Jevons (1871) and Walras (1874), economists found more and more analogies with physics, taking from physics itself ideas and the mathematical method. In fact, the central idea of *equilibrium* in economics was taken from physics (Mirowski, 1989, pp. 238-241).

In 1862 Jevons gave a speech entitled "A brief account of a general mathematical theory of political economy" at the Section F of the British Association, and Walras made clear in many occasions that he wanted to transform political economy into a hard science – the change of name of the discipline in the late 1870s from *political economy* into *economics* is not a simple coincidence. Walras' 1909 article *Économique et mécanique* synthesises the point. Irving Fisher, an important American contributor to Neoclassical theory, devoted the third chapter of his doctoral thesis – mathematical investigations in the theory of value and prices – to "mechanical analogies", providing a table of correspondences between mechanics and economics, so that a 'particle' in mechanics corresponds to an 'individual' in economics, 'energy' to 'utility' and so on (Fisher, 1892, pp. 85-86). Worthy of a comment is the fact that Fisher's thesis was reprinted in book form twice, in 1925 and 1926. Thus, inexorably, economics became a mathematical discipline, and as a rule, economists had – and, even more today, have – to be trained in mathematics and statistics.

More improvements on the use of mathematical tools and simulations within economics come from the evolutionary school of thought, within which complexity is structurally acknowledged and studied. A symbolic passage is the publication of the book *An evolutionary theory of economic change* (Nelson and Winter, 1982) in which, incidentally, the Schumpeterian model is thoroughly discussed (see pp. 289-301 therein).

Paraphrasing the Nobel-prize winner for economics Gérard Debreu, one can say that theoretical physics was the ideal toward which economic theory strove (Debreu, 1991) and it is somehow surprising that the noun 'econophysics' appeared as late as in the second half of the1990s, and applied only to finance (De Liso and Filatrella, 2002).

Innovation and technical change have been subjected to mathematical analysis, and different concepts, methods and tools have been elaborated. In particular, improvements in mathematical techniques, computation facilities and data availability have made it somehow easier to handle dynamic phenomena.

One peculiar form of technical change is that which has been defined 'sailing-ship effect', that is, the process whereby an old technology is improved as a new one, competing with it and potentially supplanting it, appears.

The process is represented graphically in Figure 1 hereafter. The picture is clear and simple, but the way to obtain it requires the econophysics toolbox in terms of use of a nonlinear functional map in which the Fermi-Dirac function has been adapted to model some economic effects – we shall illustrate the whole process in the next sections.

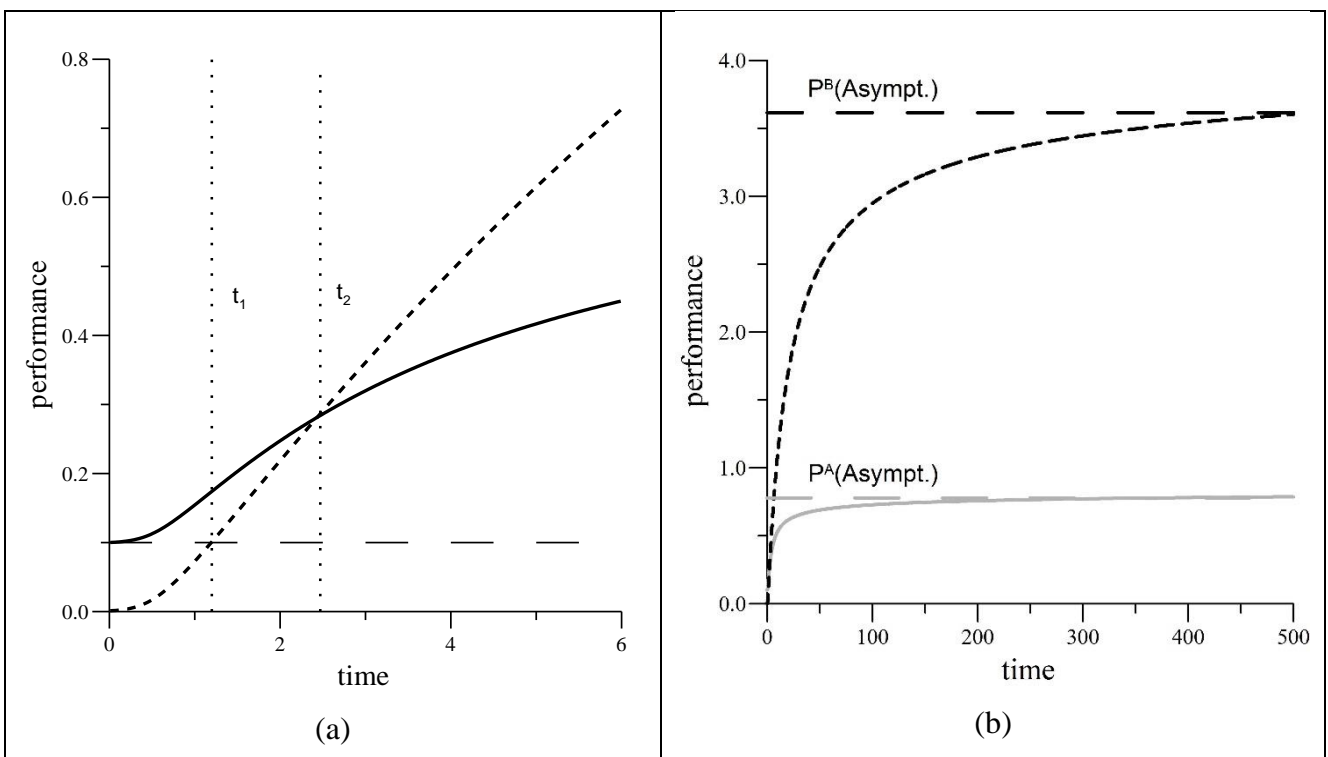

Figure 1: Qualitative behaviour of the competition between two technologies that illustrates the sailing ship effect.

Here we visualise the meaning of the phrase 'sailing-ship effect'. On the vertical axis one reads the performance of the technology – e.g. in terms of freight capacity or data transmission in MBps –, while on the horizontal axis we have time. In panel (a) of the figure the 'sailing-ship effect' is illustrated. The solid line represents the old technology, the dashed line the new technology. As the new technology appears as a niche technology with a tiny market share (Coles *et al.*, 2018) at time 0 the – by now – old technology gets improved. Had the old technology not been improved, the overtaking of the new would have taken place at time $t_1$ at a performance level 0.1. Through the sailing-ship effect, overtaking of the new over the old technology takes place later in time, at $t_2$, and at a higher performance level. The old technology gets a new lease of life and the overall technological performance is improved. In panel (b) we see the asymptotic performance of both technologies. It is important to underline that in this model the performances' dynamics are not given by an *assumed* exogenous growth ratio as, for instance, in Cantner and Vannuccini (2017).

## 3. The initial model capable of reproducing the 'sailing-ship effect'

As we mentioned in the introductory section, the starting point of this paper is the model originally proposed in De Liso and Filatrella (2008), and it is necessary here to provide a résumé of that model in order to grasp the analysis proposed in section four. However, while providing this résumé we will also provide, in sub-section 3.2, further explicit passages which allow for a deeper understanding of the original model.

### 3.1 The original model

The context is that of a dynamic duopoly race between two firms which produce a certain good or provide a certain service (e.g. data transmission) characterised by a performance level *P* (for the sake of simplicity hereafter we will always write 'product'); the higher the performance, the more consumers are attracted by that product. Firms' objective consists in profit maximization as indicated by eq. (1). The incumbent firm uses a certain 'old' technology *A*, while the new entrant firm uses a 'new' technology *B* (the words 'firm' and 'technology' will be used as synonyms, that is we will speak of 'firm *A*' or 'technology *A*' indifferently, and analogously for firm / technology *B*).

As there often occurs, at the beginning the new technology is rather unwieldy and unreliable, so that its initial performance is inferior to the old one, but characterised by a higher performance potential (e.g. cables-cum-modem vs. fiber optics in data transmission). The emergence of the new technology is a threat for the incumbent, so that the latter begins to look for ways – even in unorthodox ways (Schiavone 2013, 2014) – to improve its performance to try to keep the entrant at bay. Both technologies, though, are improved by allocating resources to R&D. Market share depends on the technical performance. The simulations are based on the following equations.

At time $t+1$ the profits $\pi$ of firm A (B) are given by:

$$\pi(t+1) = (p-c)q(t+1) - R(t)(1+r) \qquad (1)$$

where $p$ is the price of the product sold in the market; $c$ is the cost of production; $q(t)$ is the quantity of product sold; $R$ the resources devoted to R&D in order to improve the performances of the product/technology; $r$ is the rate of interest.

The latter parameter, $r$, is important in that it 'weights' the resources spent on R&D. In fact, if the firm must borrow money from banks to carry out R&D, it must pay an interest on that money, which thus increases the costs of R&D itself[1]. Money lenders may behave differently, according to the risk which they attach to different customer-firms. When firms ask for loans, banks may apply a lower interest rate to a firm with which they have long-lasting relationship, or may consider the old technology less risky to improve; should this be the case we would have $r_A < r_B$. In figure 4 and 5 of section four we will consider the different possibilities concerning the rate of interest as applied to the two firms.

If there is only one monopolistic firm in the market $R$ equals zero (there is no need to spend money to improve the product, the performance of which remains thus constant), so that $q$ corresponds at first to the total market $Q$.

When the new firm / technology emerges, we will have a market share for A and B, i.e. $S_A = q_A/Q$ and, analogously, $S_B = q_B/Q$. Consumers choose the product on the basis of its performance so that when the two technologies have the same technical performance, market share will be equally divided at 50% for each firm. Price and cost are constant, so that competition concerns performance of the technology, that is R&D is performance-enhancing rather than cost-reducing.

Performance, $P$, of the technology improves according to the following equation:

$$P(t+1) = P(t) + R(t)\gamma \exp\left(-\frac{P^M}{P^M - P(t)}\right) \qquad (2)$$

In eq. (2) we know, from the previous comments, that $R$ are the resourced devoted to R&D, and we must explain $\gamma$ and the exponential function. The former is a parameter which converts R&D expenditure into improved performance, while the latter is a function capable of limiting the efficiency of R&D expenditure when performances approach the limit value $P^M$. The latter comment refers to the fact that the closer a technology gets to its achievable maximum performance, the more it is difficult to improve it: that is, equal amounts of resources devoted to R&D generate smaller and smaller performance improvements.

---

[1] Should the firms carry out R&D with their own resources, they would miss the interest rate that they would have received, had those same resources been invested in financial assets.

In detail, eq. (2) works as follows. *For P(t)<<P^M*, the exponential reads $e^{-1}$, and therefore each money unit is converted into $\gamma/e \approx 0.37$ improvement of performances. As performances increase, it is convenient to Taylor-expand eq. (2):

$$exp\left(-\frac{P^M}{P^M-P(t)}\right) = \sum_{k=0}^{\infty} \frac{1}{k!}\left[\frac{-P^M}{P^M-P}\right]^k = \frac{1}{e}\left[1 - \frac{P}{1-P^M/e} - \cdots\right] \quad (3)$$

Therefore, for small values of the performances *P* the function decreases linearly, to mimic the decreasing returns of R&D investments. As *P* approaches the limit value $P^M$ the function (2) exponentially decreases to 0, that is any investment in R&D results in a negligible increase of the performances.

The competition process is based on profit maximization of both firms. Profit depends on the market share *S* which, in turn, depends on one's own performance but also on the performance of the other technology – that is $S_A(t) = f_A (P_A(t), P_B(t))$ and $S_B(t) = f_B (P_B(t), P_A(t))$ – while one's performance improves according to the resources devoted to R&D, and is limited by the highest limit set by some physical law.

Functions $f_A$ and $f_B$ have the following properties, as dictated by economic reasoning: a) they have the same functional shape (no *a priori* preference for one technology or the other); b) they only depend upon the ratio between the performances (users perceive only the relative strengths of the two technologies); c) the functional form is a sigmoid as observed in many empirical studies on technology adoption.

To keep the model as manageable as possible, the abovementioned properties can be modelled by the formula:

$$S_{A/B}(t) = f_{A/B}(t) = \left[\frac{P_{A/B}(t)}{P_A(t)+P_B(t)}\right]^{\frac{1}{\alpha}-1} \left(\frac{1}{2}\right)^{2-\frac{1}{\alpha}}. \quad (4)$$

Eq. (4) satisfies the 'natural' requirement according to which the share must be 1/2, or 50%, when the performances of the two technologies are identical. The parameter $\alpha$ – which must be $0 < \alpha < 1$ – describes the sensitivity of the consumer to the quality of the product: the higher the value, the more the consumer is indifferent, or incapable to discern, the quality of the two technologies. In the limit $\alpha \to 0$ the share approaches the step function $\vartheta\left[\frac{2P_{A/B}(t)}{P_A(t)+P_B(t)}\right]$ (the best performance gets all), and for $\alpha=1$ the share reads $S_{A/B} = 1/2$ (market is divided in half independently of the products' relative quality). For simplicity we consider here the intermediate case $\alpha=1/2$, for which $S_{A/B}(t) = \left[\frac{P_{A/B}(t)}{P_A(t)+P_B(t)}\right]$.

The equations that are simulated in the next section are thus:

$$P_A(t+1) = P_A(t) + \tfrac{1}{2}\big(P_A(t)+P_B(t)\big) \times \left[1 - (1+r_A)\frac{(P_A(t)+P_B(t))^2}{P_B(t)(p-c)Q\gamma_A} \exp\left(\frac{-P_A^M}{P_A^M - P_A(t)}\right)\right] \times$$

$$\times \vartheta \left[1 - (1+r_A)\frac{(P_A(t)+P_B(t))^2}{P_B(t)(p-c)Q\gamma_A} \exp\left(\frac{-P_A^M}{P_A^M - P_A(t)}\right)\right] \quad (5a)$$

$$P_B(t+1) = P_B(t) + \tfrac{1}{2}\big(P_A(t)+P_B(t)\big) \times \left[1 - (1+r_B)\frac{(P_A(t)+P_B(t))^2}{P_A(t)(p-c)Q\gamma_B} \exp\left(\frac{-P_B^M}{P_B^M - P_B(t)}\right)\right] \times$$

$$\times \vartheta \left[1 - (1+r_B)\frac{(P_A(t)+P_B(t))^2}{P_A(t)(p-c)Q\gamma_B} \exp\left(\frac{-P_B^M}{P_B^M - P_B(t)}\right)\right] \quad (5b)$$

Let us remind that in the above equations *p* is the unitary selling price of the product – not to be confused with performance *P* –, *c* the production cost of a unit, and *Q* the total market; thus *Q(p-c)* is the market value at stake in the competition that is considered to be constant. Should one technology – usually the new one – become much better than the other after some iterations, that technology could corner the market, i.e. virtually all users would adopt it.

The only component which remains to be explained in eq.s (5a) and (5b) is the $\vartheta$ function. This is a *step* function, and is needed because of economic logic. The battle between the two technologies is bound to have an end: in the long run, performance improvements become more and more difficult and more costly, so that further expenditure on R&D is not (more than) offset by higher receipts. Thus, when the argument within the last brackets turns negative, R&D expenditure must fall to zero – and this is what the $\vartheta$ function does.

### 3.2 Analytical considerations and some more clarifications on the working of the model

Most of the information that can be retrieved from the model stems from simulations – as anticipated in Fig. 1 – and the two-dimensional nonlinear map (5a) and (5b) is sufficiently simple to allow for some analytical considerations. However, we need to elaborate the original model in order to make explicit some passages which were left unspoken there. First of all, it is convenient to rewrite eq.s (5a,b) as a formal two-dimensional map:

$$\begin{cases} P_A(t+1) = G_A(P_A(t), P_B(t); P_A^M, r_A, \gamma_A, (p-c), Q) & (6a) \\ P_B(t+1) = G_B(P_A(t), P_B(t); P_B^M, r_B, \gamma_B, (p-c), Q) & (6b) \end{cases}$$

Then we can identify the asymptotic solution, i.e. the fixed point. In fact, when iterations do not lead to any further change, the fixed point, that is $P_{A/B}(t+1) = P_{A/B}(t) = P_{A/B}^*$, reads:

$$\begin{cases} P_B^* = (1 + r_A) \dfrac{(P_A^* + P_B^*)^2}{(p - c)Q\gamma_A} exp\left(\dfrac{-P_A^M}{P_A^M - P_A^*}\right) & (7a) \\ P_A^* = (1 + r_B) \dfrac{(P_A^* + P_B^*)^2}{(p - c)Q\gamma_B} exp\left(\dfrac{-P_B^M}{P_B^M - P_B^*}\right) & (7b) \end{cases}$$

The fixed point of the map, that in economic terms represents the equilibrium point at which the two technologies do not improve any further, is only implicitly defined by the above equations. From the mathematical point of view the $\vartheta$ function seen above introduces a discontinuity that makes the map non-differentiable precisely in the fixed point, for the eigenvalues of the Jacobian that regulates the stability of the map in the fixed point (7a, 7b) are not defined:

$$det \begin{pmatrix} \dfrac{\partial G_A(P_A, P_B)}{\partial P_A} - \lambda & \dfrac{\partial G_A(P_A, P_B)}{\partial P_B} \\ \dfrac{\partial G_B(P_A, P_B)}{\partial P_A} & \dfrac{\partial G_B(P_A, P_B)}{\partial P_B} - \lambda \end{pmatrix} \Bigg|_{\substack{P_A = P_A^* \\ P_B = P_B^*}} \quad (8)$$

Thus, the stability of the fix point is to be established numerically. Another interesting property is the limit in which the technology $A$ (or $B$) does not invest. This happens when the $\vartheta$ function argument is negative, that is, for instance for technology $A$:

$$P_A \leq (1 + r_B) \dfrac{(P_A + P_B)^2}{(p - c)Q\gamma_B} exp\left(\dfrac{-P_B^M}{P_B^M - P_B}\right) \quad (9a)$$

Condition (9a) can be seen as the maximum value that can be reached by performances of the old technology. It is interesting to notice that performances are squared on the right hand side of the equation, while they are 'only' linear at the left hand side. Thus, for any assumed value of the performances $B$, eventually inequality (9a) holds true, and $A$'s expenditure in R&D ceases. An analogous expression can be found for technology $B$:

$$P_B \leq (1 + r_A) \dfrac{(P_A + P_B)^2}{(p - c)Q\gamma_A} exp\left(\dfrac{-P_A^M}{P_A^M - P_A}\right). \quad (9b)$$

Eq. (9b) describes the maximum performances actually achieved by the new technology, $B$. For any value of $A$, this limit performance is below the maximum theoretical value $P^M{}_B$. Finally, the condition (9b) is clearly true for arbitrary small values of $B$'s performances. In economic terms, the new technology $B$ invests, no matter how small are the initial performances – or how high are the performances of the old technology.

# 4. Simulations of the model

How the parameters affect the race for technological primacy has been investigated through extended simulations of the model. The advantage of a quantitative model lies precisely in the possibility to compute the consequences of changes of external conditions (e.g. interest rate) or to evaluate the role of intrinsic features of the technologies (i.e. the maximum performance which can be reached by the technology and the efficiency in converting resources devoted to R&D into improved performance of that technology).

In particular, we will study what happens when we deal with a changing 'economic' parameter, i.e. the interest rate $r$, on the one hand, and how 'scientific' and 'technological' parameters, i.e. $\gamma$ and $P^M$, affect the competition process between the two technologies, on the other. Let us begin with the latter.

## 4.1 'Scientific' and 'technological' parameters

In Figure 2 we compare what happens when we keep constant the old technology's capability of converting R&D into improved performance, i.e. $\gamma_A = 2.55$, while we have different values of $\gamma_B$, the other parameters being $P^M_B = 5$, $P^M_A = 1$ (i.e. the new technology has a potential five times bigger than the old) and $r = 5\%$ for both technologies.

We plot the asymptotic value of the performances achieved by the old technology $P_A$ and the new technology $P_B$. In the language of the nonlinear maps, we numerically retrieve the fixed points of eqs (7a and 7b). It is important to stress the fact that in the figure we do not plot the time evolution, but the limit value of either technology when $\gamma_A$ is held constant while $\gamma_B$ takes different values. In the numerical simulations we have never encountered the case of an unstable fix point, e.g. encountering a bifurcation or a more complex orbit.

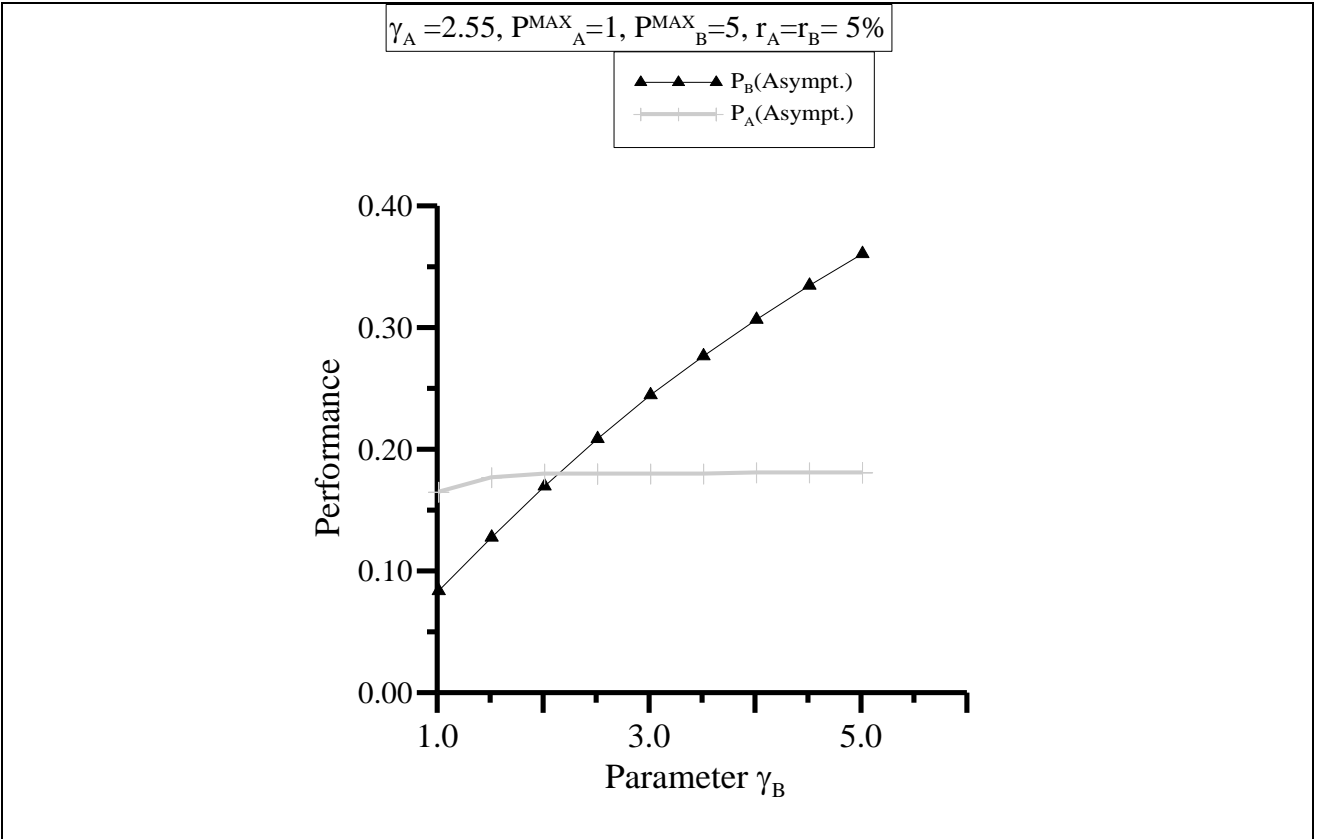

Figure 2: effect of the efficiency of the conversion of R&D investments on the asymptotic (final) performances of the two technologies.

As one would expect, the larger the efficiency of the conversion of R&D into improved performance of the new technology the higher the asymptotic performance level. However, the efficiency term γ seems to affect also the qualitative outcome: in fact, for low efficiency values, the old technology is capable to withstand the competition and to maintain the primacy in the long run – see grey line in Fig. 2. Only for values of $\gamma_B$ above 2 is the new technology actually capable to impose itself as the leading technology. Efficiency is also capable to affect the race in a subtler way: an increase in the efficiency of the new technology – and this is true up to $\gamma_B = 2$ – stimulates improvements of the old technology, which thus achieves higher levels of (asymptotic) performance.

For sake of simplicity we have shown here the dynamics of the performances. As the share, or penetration, of the product in our model is determined by the performances through the functions $f_A$ and $f_B$ – see Section 3.1 –, the dynamics of the performances translates into an evolution of the market shares that can be more directly related to the empirical data (see, for instance, Palacios-Fenech and Tellis, 2016).

The role played by the maximum performance of the new technology, $P^M_B$, is investigated in Figure 3. As we said, the maximum performance of a technology is limited by a physical law. Setting the parameters $\gamma_A = \gamma_B = 2.55$, $P^M_A = 1$ and $r = 5\%$ for both technologies, we study what happens when $P^M_B$ can take different values.

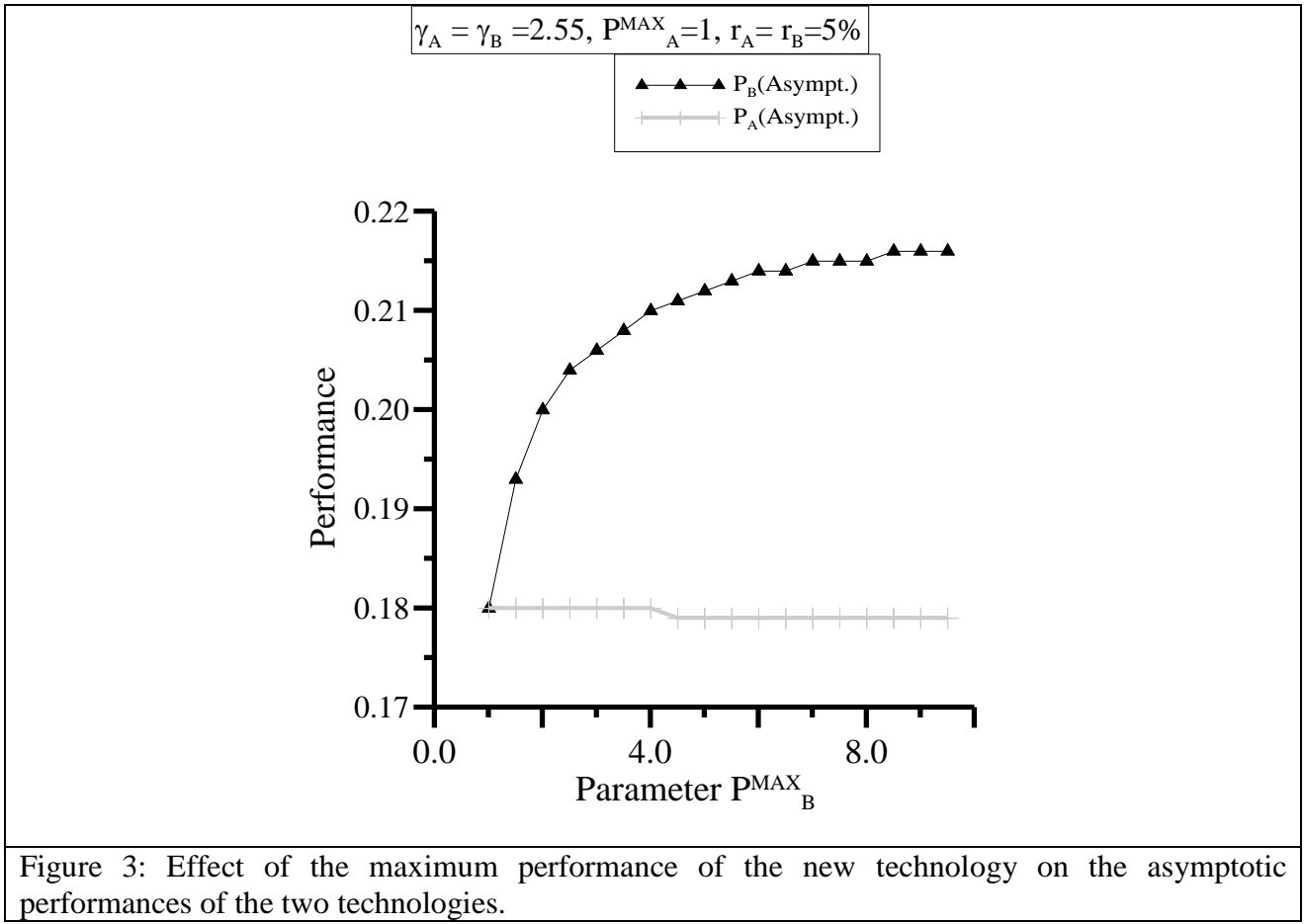

Figure 3: Effect of the maximum performance of the new technology on the asymptotic performances of the two technologies.

First of all, as one would expect, the two technologies reach the same performance level when both technologies have the same potential. Then, as the potential of the new technology increases, the incumbent technology even experiences a diminution in its technical performance. Once more, let us point out that this is not the time evolution of the performance of either technology, but the behaviour of the performance of technology A when different values of the maximum performance of technology B are considered. Thus, values of $P^M_B$ greater than 4 actually create a condition in which technology A's performance deteriorates.

*4.2 The economic parameter*

An interesting aspect of the simulation is that which considers the role played by the economic parameter *r*, that is the interest rate. If firms borrow money from banks to carry out R&D, then we are in the Schumpeter 'Mark I' condition[2], so that, in fact, the ultimate decision-maker to affect the innovative process is the bank.

A first simulation, whose results are illustrated in Figure 4, considers the situation in which both technologies are asked to repay the same interest rate. Setting γ at the same value for both technologies ($\gamma_{A,B} = 2.55$) and the new technology five times potentially better than the old one ($P^M_B = 5$, $P^M_A = 1$) one can see that the higher the interest rate, the lower the performance level reached by both technologies.

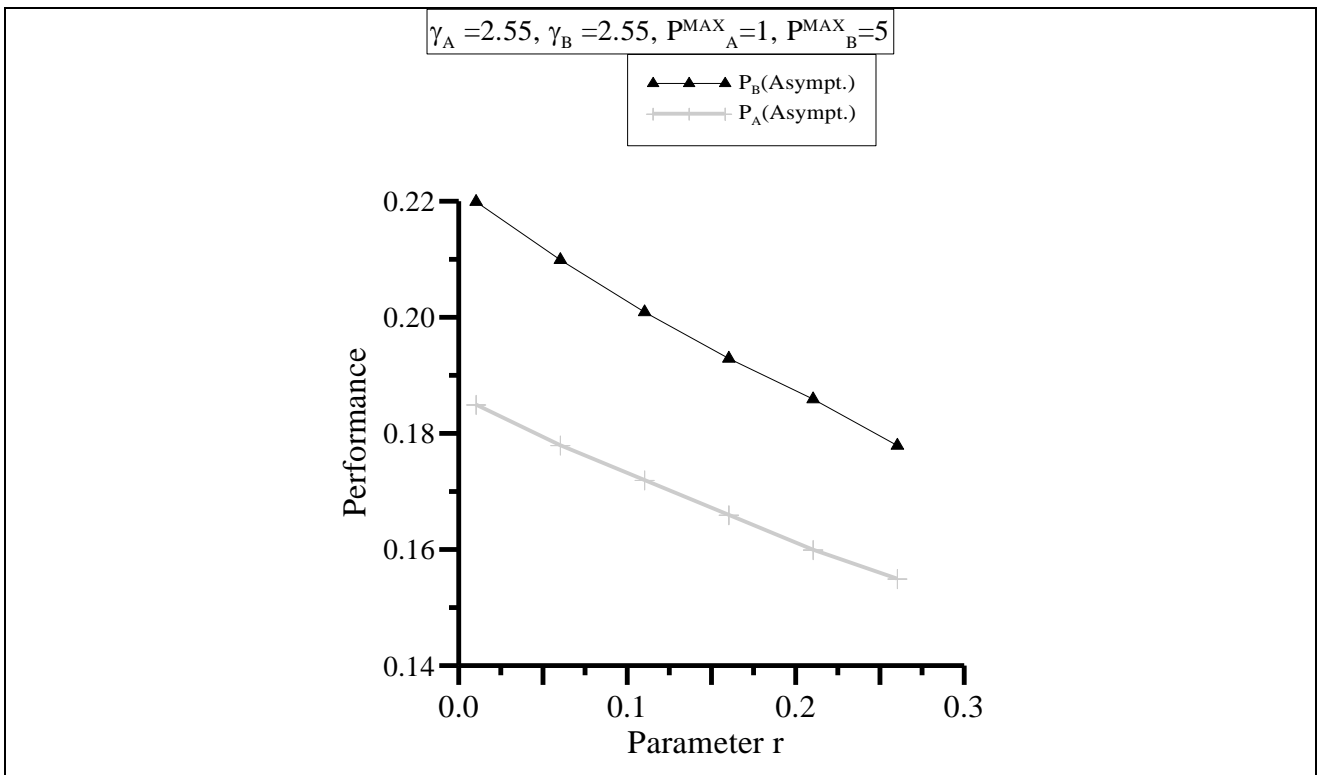

Figure 4: effect of the interest rate on the asymptotic performances of the two technologies; here the interest rate *r* applied to both firms / technologies is the same.

When the interest rate is uniformly applied to both technologies, the new technology, characterised by a higher potential performance, always prevails.

In Figure 5, instead, we see what happens when the old technology *A* borrows money for R&D at a constant interest rate $r_A$, while technology's *B* varies, the other parameters being $\gamma_A = \gamma_B = 2.55$, $P^M_A = 1$ and $P^M_B = 5$. In panel (a) of the figure the interest rate for technology *A* is 10%. Obviously, the lower the interest rate for *B*, the higher its performance. Technology *A* and *B* have the same technical performance when *A* pays 10% and *B* 30%, thus for $r_B$< 30% B's performance is higher than A's.

---

[2] Joseph Schumpeter writes that it is the banking system which makes innovation possible by risking the capital, defining the banker as the ephor of the exchange economy (1911, English translation 1961, p. 74). Despite the fact that big corporations very often do not depend on the banking system because they have their internal resources (this would be the 'Schumpeter Mark II' situation), new innovative start-ups actually *do* often depend on that banking system. The latter, by selecting which projects to finance, applying different interest rates, implicitly affects the innovative path.

The competition process will experience the prevalence of one technology or the other according to the interplay between the value(s) taken by the interest rate on the one hand, and the potential maximum performances on the other. In particular, the higher the interest rate to borrow money, the less one will invest in R&D. Put it in another way, when the interest rate is excessively high, technology *B* stops the developments too early, and in spite of the fact that it is definitely better – potentially five times – than technology *A*, the actual performance which it reaches is much lower. Specifically, for the parameters of Fig. 5, should the interest rate for *B* be higher than 30%, technology *B* would remain inferior to technology *A* for lack of research. Instead, when investments are affordable, e.g. when firm *A* and *B* have the same interest rate of about 10%, *B*'s asymptotic performance is $P_B$=0.203 while *A*'s is $P_A$=0.173.

In panel (b) of Figure 5, *A* pays a 5% interest rate, the other parameters being the same as in panel (a). First of all, one sees that a lower interest rate for technology *A* implies a better performance for *A* itself with respect to what happens in panel (a). In panel (b), technology *B* has the same performance as technology *A* when firm *B* gets a 25% interest rate. As soon as *B* pays an interest rate smaller than 25%, *B* itself overtakes *A* in terms of performance. Once more, this phenomenon is explained by the fact that the higher interest rate is offset, or more than offset, by a much higher maximum potential performance, which makes it easier to transform resources devoted to R&D into improved performance. When the two firms are charged the same interest rate, i.e. 5% in this case, *A*'s performance is 0.179, while *B*'s is 0.212.

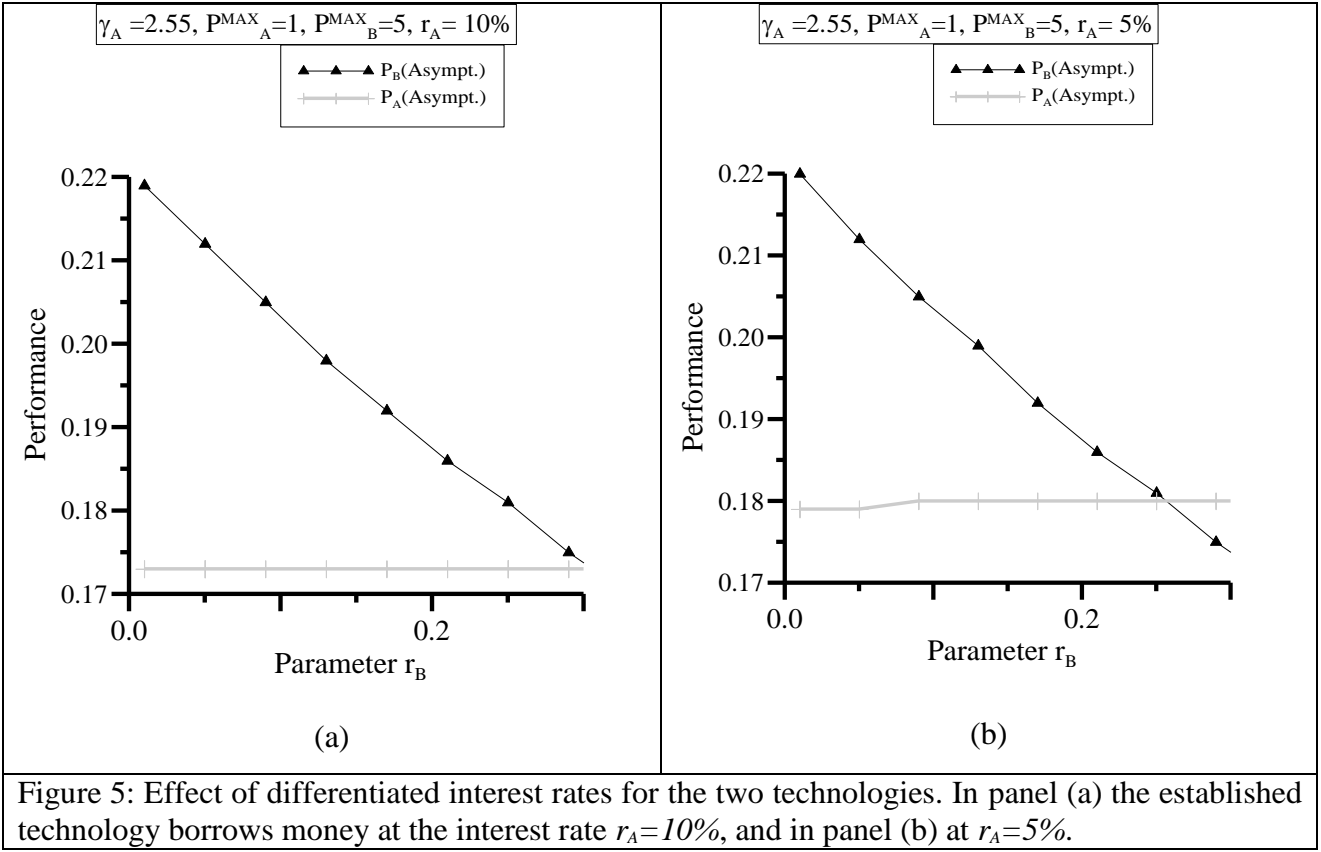

Figure 5: Effect of differentiated interest rates for the two technologies. In panel (a) the established technology borrows money at the interest rate $r_A=10\%$, and in panel (b) at $r_A=5\%$.

## 5. Conclusions

Technology and technological change are at centre stage of our societies at least since the inception of the First Industrial Revolution. Within technology scientific, 'pure' technological and economic dimensions merge, so that when we want to study technological evolution, complexity is a structural feature which we have to face.

In this work we have focused our attention of that form of technological change which is known as the 'sailing-ship effect', that is those improvements in an old technology which are intentionally sought as a new, potential supplanter, technology emerges.

The dynamic duopolistic competition between the new and the old technology is studied through a nonlinear functional map, which provides a quantitative tool to examine the evolution of the performance of both technologies through time. The overall process is based on firms' profit maximization; firms compete for market share, and the latter depends on technologies' performance – i.e. the better the technical performance, the higher the market share.

The outcome of this competition, though, depends on the three key parameters which have been considered, namely $\gamma$, $P^M$ and $r$. As we pointed out in section three, the first is scientific-technological, the second is purely technological and the third is economic in kind.

Parameter $\gamma$ synthesises the way in which resources devoted to R&D can be translated into higher performance of the technology. Obviously, the higher is $\gamma$, the higher the performance becomes, and it is interesting to see – as we did in Figure 2 – what happens when the old technology evolves according to a given value of $\gamma_A$, while different values of $\gamma_B$ are considered.

Parameter $P^M$ is the physical upper technical limit of the technology. As a rule, the new technology will be characterised by a higher potential with respect to the old, so that, given this premise, the new technology has always an advantage over the old.

The interest rate $r$ plays a key role, and the higher it is, the lower the performance of both technologies.

What is interesting is the result of the system's behaviour when different values of the three parameters are considered, and in particular the role played by $r$. In fact, as we emphasised, this is an economic parameter which heavily affects the overall system and it is a purely discretionary parameter, chosen by the banking system. The rate at which the bank makes money available to firms can make a big difference in terms of which technology succeeds, so that the bank becomes, in the end, a key agent in promoting or withholding the old or the new technology.

The model here used can thus be useful for scenario studies in which an old and a new technology compete to conquer a market. The multidimensional parameters space can be thoroughly explored with extensive simulations, that can highlight how the technological and economic properties are quantitatively interwoven.

The comments above open the way to policy issues. Firms – and governments – have to decide if and how much to invest to improve technologies, and the first decision that the owners / users of the old technology have to take is whether the battle with new technology is worth making. Decisions are largely based on scientific, technological and economic *expectations* concerning one's own technology *and* the other. A complex picture emerges and, broadening for a moment the perspective, one should take into account that nowadays the financing of innovation is more and more characterised by the presence of venture capitalists and of the so-called 'business angels'. Both latter categories may have a risk-neutral or even a risk-loving attitude, while banks tend to be risk-averse, so that those who deal with the former agents have an advantage over those who have to refer to the traditional banking system to finance their innovation efforts.

In closing, let us hint at possible extensions of the model. In the first place, the present model describes competition between two firms (or technologies). One could try to explore a situation in which competition takes place between more than two technologies; a challenge would thus be to extend the exercise to the case in which more two agents exist, possibly making also use of game theory.

Secondly, when we deal with R&D, the transformation of resources devoted to R&D itself into improved performance may be affected by the presence of a stochastic, or 'noisy', component. With regards to the role of noise in a competition dynamics, inspiration could be found in the work by Giuffrida et al. (2009) or Valenti et al. (2016). One could for instance try to consider a stochastic component in the sailing-ship competition process – a component which would be strictly connected, e.g. in a multiplicative way, with the way in which our parameter γ works.

**Compliance with Ethical Standards**

Authors declare that they have no conflict of interest.

Ethical approval: This article does not contain any studies with human participants or animals performed by any of the authors.